\documentclass[epj]{svjour} 
\usepackage{amssymb}
\input  epsf 

\begin{document}

\title{Stretched exponential relaxation in the Coulomb glass}

\author{A.\ D\'\i az-S\'anchez \inst{1,2} 
\thanks{\emph{Present address:} andiaz@upct.es}
\and A. P\'erez-Garrido \inst{1}
}
\institute{
Departamento de F\'\i sica Aplicada,
Universidad Polit\'ecnica de Cartagena, \\
Campus Muralla del Mar, Cartagena, 
E-30202 Murcia, Spain.
\and 
Dipartamento di Scienze Fisiche, Universit\'a di Napoli
``Federico II'', \\ 
Complesso Universitario di Monte Sant'Angelo, 
Via Cintia , I-80126 Napoli, Italy \\
and INFM, Unit\'a di Napoli, Napoli, Italy.}

\date{Received: \today / Revised version: \today}

\abstract{
The relaxation of the specific heat and the entropy to their equilibrium values is 
investigated numerically for the three-dimensional Coulomb glass at very low temperatures. 
The long time relaxation follows a stretched exponential function, 
$f(t)=f_0\exp\left[-(t/\tau)^\beta\right]$, with the exponent $\beta$ 
increasing with the temperature. The relaxation time follows an Arrhenius
behavior divergence when $T\rightarrow 0$. A relation between the specific heat 
and the entropy in the long time regime is found.}


\maketitle

\section{Introduction}

The relaxation in most glassy systems such as structural glasses, ionic
conductor, supercooled liquids, polymer, colloid, and spin glasses \cite{EJ1}
deviates strongly from a single exponential relaxation at some temperature
$T^*$ well above the static transition. In the three dimensional Ising spin 
glass model Ogielski \cite{O85} observed that the long time regime of the 
relaxation functions is well approximated by the non-exponential function 
\begin{equation}
f(t)=f_0t^{-x}\exp\left[-(t/\tau)^\beta\right]\,,
\end{equation}
below $T^*=T_G$, where $T_G$ is the Griffiths temperature \cite{G69}. 
In supercooled liquids \cite{GS92} the characteristic features of the relaxation 
processes in the long time regime is captured by a stretched exponential or 
Kohlrausch-Williams-Watts \cite{KWW54} decay function
\begin{equation}
f(t)=f_0\exp\left[-(t/\tau)^\beta\right]\,,
\end{equation}
with $0<\beta\le 1$. This behavior can be obtained by a superposition of 
purely exponential relaxation process but subject to a broad distribution of 
relaxation times \cite{RB94}.
 
The Coulomb glass \cite{PO85,SE84} is a prominent example of 
glassy systems. In heavily doped crystalline semiconductors,
amorphous semiconductor-metal alloys, and granular metals,
it plays an important role as a semiclassical model for systems of
localized states. The out of equilibrium dynamical behavior of the 
Coulomb glass has been studied previously by Schreiber et al.\ 
\cite{ST94} and P\'erez-Garrido et al.\ \cite{PO98}. They 
determined numerically the transition probabilities between low-energy 
many-particle states, and studied the eigenvalues of the transition probability 
matrix. A broad distribution of relaxation times over 
several orders of magnitude was found in both cases, which reflects the glassy 
behavior of this system. This broad distribution bring to a power law relaxation 
function \cite{PO98}. Moreover, Wappler et al.\ \cite{WS97} used the 
damage-spreading algorithm to study the temporal evolution of the system in thermal
equilibrium, and found evidence for a dynamical phase transition at a temperature
$T^*$ which depended of the degree of disorder considered. Yu studied the time 
development of the Coulomb gap considering a 
self-consistent equation for the density of states \cite{Y99}. She also observed 
that very long time scales are involved. D\'\i az-S\'anchez et al.\ \cite{DM98,DM99}, 
considering the dynamic in the configuration space, studied non-ergodic effects on 
the specific heat and found very long relaxation times.
Pastor et al.\ \cite{PD99} predicted the existence of a glass
phase for a mean-field model of interacting spinless fermions in the
presence of disorder. Experimental measurement in electronic system of an Anderson
insulator \cite{VO00} showed a glass phase with aging phenomenon appearing. There it
was found that the common relaxation law can be fitted by a stretched exponential 
function.

The aim of this paper is to study the shape of the relaxation process of 
the specific heat and the entropy in the Coulomb glass at very low temperatures and 
in thermal equilibrium. The paper is organized as follows: Section 2 introduces 
the Coulomb glass model. Section 3 describes the numerical procedure, 
i.e. how the low-energy many-particle states are obtained numerically, and the 
dynamic used in the simulation. In Section 4, we use this procedure for the 
investigation of the relaxation process of the specific heat and the entropy. 
Finally, in Section 5 we extract some conclusions.

\section{The model}

A practical model to represent the Coulomb glass problem for simulating 
an impurity band of localized states in lightly doped semiconductors, 
when quantum interference can be neglected, was proposed in references  
\cite{PO85,SE84}. Later it was also applied to simulate granular metals 
\cite{PA92} and conducting polymers \cite{LC93}. Following 
reference \cite{BE79}, we consider a three dimensional cubic lattice with  
$N$ sites. The sites can be occupied by $KN$ ($0<K<1$) electrons which
are interacting via an unscreened Coulomb potential. Background 
charges $-K$ are added at each of the lattice sites, guaranteeing 
electro-neutrality. The disorder is simulated by a random potential $\epsilon_i$. 
Their values are uniformly distributed between $-B/2$ and $B/2$. This model 
is represented by the Hamiltonian
\begin{equation}
H=\sum_i(\epsilon_i-\mu) n_i +\sum_{i<j} {\frac{(n_i-K)(n_j-K)}
{r_{ij}}}\,,
\end{equation}
where $n_i \in \{0,1\}$ denotes the occupation number of site
$i$, $r_{ij}$ is the distance between sites $i$ and $j$ according 
to periodic boundary conditions \cite{MR53}, and $\mu$ is the chemical
potential. The lattice spacing is taken as unit of distance. 

In this paper, we focus in the half-filled impurity band $K=1/2$, where 
there is particle-hole symmetry and the chemical potential
$\mu$ equals zero at any temperature in the thermodynamic limit. We take
$B=2$ in this paper, similar results to the ones presented here
are found for other values of $B$ \ \cite{DM99}.

\section{Numerical procedure}

Our aim is to simulate the temporal evolution of the Coulomb 
glass at very low temperatures in thermal equilibrium. For that we first obtain 
a set ${\cal S}$ of low-energy many-particle configurations and later define 
a dynamic between these configurations. We apply this procedure to study 
the relaxation of the specific heat and the entropy to their equilibrium values.
  
\subsection{Low-energy configurations}

We find the low-energy many-particle configurations by means of a
three-steps algorithm \cite{DM99}. This comprises local search
\cite{PO97,MP96}, thermal cycling \cite{MN97}, and construction
of ``neighboring'' states by local rearrangements of the charges
\cite{PO97,MP96}. The efficiency of this algorithm is illustrated in
reference \cite{DM99}. In the first step we create an initial set
${\cal S}$ of metastable states. We start from states chosen at random
and relax these states by a local search algorithm which ensures
stability with respect to excitations from one up to four sites. In the
second step this set ${\cal S}$ is improved by means of the thermal
cycling method, which combines the Metropolis and local search
algorithms. The third step completes the set ${\cal S}$ by
systematically investigating the surroundings of the states previously
found.  

\subsection{Dynamic in the configuration space}

We now present a method to study the influence of the duration of the measurement, 
i.e. the observation time $\tau_{\rm m}$, on the expectation value of the specific 
heat $c$ and the entropy $S$ \cite{DM99}. This method defines a dynamic in the 
configuration space. 

During $\tau_{\rm m}$ the state of the sample travels randomly through its 
configuration space. A transition between configurations $I$ and $J$ is allowed if the 
transition time $\tau_{IJ}$ is shorter than $\tau_{\rm m}$. So, at finite $\tau_{\rm m}$ 
only transitions with characteristic time $\tau_{IJ}$ shorter than $\tau_{\rm m}$
contribute to the specific heat. Thus, for a given $\tau_{\rm m}$, we consider two
configurations as connected if their $\tau_{IJ}$ is shorter than $\tau_{\rm m}$,
and we group the configurations in clusters according to these connections.
These clusters correspond to regions of the configurations space being isolated
from each other on the time scale $\tau_{\rm m}$.

The characteristic transition time between configurations $I$ and $J$ is
\cite{PO85,DM99},
\begin {equation}
\tau_{IJ}=\tau_0 \exp \left( 2\sum r_{ij}/a \right)
  \exp \left( E_{IJ}/T \right)/Z \, .
\end{equation}
In this equation, the quantity $\tau_0$ is a constant of the order of the
inverse of the phonon frequency, $\tau_0\sim 10^{-13}\ {\rm s}$. The sum is
the minimized sum over all hopping distances between sites which change
their occupation in the transition $I \rightarrow J$. $a$ denotes the
localization radius, $E_{IJ} = \max (E_I,E_J)$ where $E_I$ is the
energy of the state $I$, $T$ is the temperature, and $Z$ is the partition 
function.

Provided, at the beginning of the measuring process, the sample is
in one of the states of the cluster $\alpha$, we measure as value of
the mean energy in this cluster,
\begin{equation}
\langle E \rangle_\alpha =  \sum_{I \in \alpha} E_I
P_I\, ,
\end{equation}
with
\begin{equation} 
P_I=\exp \left(\frac{-E_I}{T} \right)\, Z_\alpha^{-1}\, ,
\end{equation}
where $Z_\alpha$ denotes the value of the the partition function of this cluster. 
The entropy of the cluster is given by
\begin{equation}
S_\alpha =  \sum_{I \in \alpha} 
P_I \ln \left(P_I \right)\,. 
\end{equation}
    
Assuming that the clusters are in thermal equilibrium the specific heat for a cluster 
$\alpha$ of states is:
\begin{equation}
c_\alpha =
\frac{\langle E^2 \rangle_\alpha - \langle E \rangle_\alpha^2}
{T^2 N} \,.
\end{equation}

The expectation value of $c$, $\langle c \rangle$, that is 
the mean value of repeated measurements, is given by the weighted average 
of $c_\alpha$:
\begin{equation}
\left< c (T,\tau_{\rm m})\right> =
\sum_\alpha c_\alpha \, P_\alpha \,,
\end{equation} 
where $P_\alpha$ is the probability to find the sample in one of the states 
belonging to the cluster $\alpha$. 
Similarly the expectation entropy,$\langle S \rangle$, is given by
\begin{equation}
\left< S (T,\tau_{\rm m})\right> =
\sum_\alpha S_\alpha \, P_\alpha \,.
\end{equation} 

Thus $\langle c \rangle$ and $\langle S \rangle$ 
depend via the cluster structure and $P_\alpha$ on $\tau_{\rm m}$, and also 
on $T$. Both the cluster structure and $P_\alpha$ are influenced by the
history of the sample. In our numerical experiments, we presume that 
the sample has reached thermal equilibrium before the measurements; 
thus $P_\alpha = Z_\alpha /Z$.

\section{Relaxation functions}

We study the process of relaxation of the specific heat and the entropy according to the
methods presented in the previous section. To make the influence of $\tau_{\rm m}$ 
directly visible, we consider the rate of the values of the specific heat and the entropy 
for finite and infinite (equilibrium) duration of measurement, respectively:
\begin{equation}
q_c(T,\tau_{\rm m}) = \frac
{\left< c (T,\tau_{\rm m})\right>}{\left< c (T,\infty)\right>}=
\frac {\left< c (T,\tau_{\rm m})\right>}{c_{\rm eq}}\,,
\end{equation}
\begin{equation}
q_s(T,\tau_{\rm m}) = \frac
{\left< S (T,\tau_{\rm m})\right>}{\left< S (T,\infty)\right>}=
\frac {\left< S (T,\tau_{\rm m})\right>}{S_{\rm eq}}\,,
\end{equation}
where $c_{\rm eq}$ and $S_{\rm eq}$ are the equilibrium values.
Moreover, in order to characterize the shape of the relaxation process in the
long time regime we study the following functions, 
\begin{equation}
R_c(T,\tau_{\rm m})=1-q_c(T,\tau_{\rm m}) \,,
\end{equation}
\begin{equation}
R_s(T,\tau_{\rm m})=1-q_s(T,\tau_{\rm m}) \,,
\end{equation}
which go to zero at the equilibrium.               

In our simulations we consider the localization radius $a$ equals always 
0.2 \cite{MP91b}. Investigating the physical properties of macroscopic systems, we have
calculated ensemble averages, and have compared the results for different
sample sizes. In order to ensure the convergence of the results on the number of 
low-energy many-particle configurations in the set ${\cal S}$, we have taken 
from $25000$ to $75000$ configuration into account for the cases studied here 
(the number of configurations to consider depends on the temperature and system size). 
This election ensures that the width of the related energy interval exceeds the
temperature by at least a factor of 25.
As in reference \cite{DM99} for $N\gtrsim 512$ the result are free of finite-size effects.
Here we present the simulations for $N=512$ taking into account 1000 samples 
for ensemble averages. In order to simplify the notation we consider the 
variable $t=\tau_{\rm m}/\tau_0$, i.e. our unit of time is $\tau_0$.

We first study the shape of the relaxation functions $R_c$ and $R_s$.
Figure 1 shows $R_c$ and $R_s$ as a function of $t$, for five different temperatures.
Here we can see that the long time regime, $R_c\lesssim 0.5$ and $R_s\lesssim 0.5$, 
may be reasonably fitted by a stretched 
exponential function. In the worse case, $T=0.014$, the $75 \%$ of the stretched 
exponential function is compared with the data although for $T=0.006$ it becomes 
the $80 \%$. For all temperatures considered here $R_c$ is equal to $R_s$
in the long time regime (within the errors).
This form of the relaxation functions brings us to think in a broad distribution 
of relaxation times in the system, as it has been found in references \cite{ST94,PO98}. 
We have also attempted a fit of our results with equation (1) but the better fit is made 
with $x\simeq 0$, i.e. we recover equation (2). 

The influence of the temperature on the relaxation process is studied.
The values of $\beta$, obtained by fitting $R_c$ and $R_s$ with equation (2), are presented 
in Figure 2. We see that $\beta$ increases with the temperature and it is the same for
both relaxation functions, $R_c$ and $R_s$, at each temperature (within the errors). 
This behavior of $\beta$ is also found in other glassy systems 
\cite{FC99}. We could expect some temperature $T^*$, higher than the temperatures 
studied here, where $\beta = 1$, recovering the exponential relaxation process at $T>T^*$.
Unfortunately our method does not let to investigate higher temperatures because we
must take a finite number of configurations into account in the set ${\cal S}$ (we have
taken the maximum of configurations that we have been able in order to make the 
calculations in a reasonable time).

For an estimate of the long time relaxation we use $\tau$ from the fit of 
equation (2) of $R_c$ and $R_s$. As we can see in Figure 3, $\tau_r$ is reasonably 
fitted by an Arrhenius behavior divergence as $T\rightarrow 0$
\begin{equation}
\tau_r(T)=a\exp[T_0/T] \,,
\end{equation} 
with $a=1.4\times 10^8$ and $T_0=0.1106$ for $R_c$; for $R_s$ we have $a=1.2\times 10^9$ 
and $T_0=0.0946$. The differences in $a$ and $T_0$ from both relaxation functions
are because of the sensitivity of $\tau$ on the details of the fit. 

We now consider the relation between the specific heat and the entropy
in the long time regime. At the equilibrium both are connected by the relation
\begin{equation}
\frac{\partial \langle S \rangle}{\partial T}=\frac{N}{T}\langle c \rangle \,.
\end{equation}                                                                                                  
As we will see below due to the temperature dependence of $R_c$ and $R_s$ this 
thermodynamic relation only hold at the equilibrium, i.e.  when $t \gg \tau_r$. 
We have seen that $R_c \approx R_s=R$ in the long time regime, so that 
$q_c \approx q_s=q$ and we can write from equations (11) and (12) 
\begin{equation}
\langle c \rangle = q(T,t)c_{\rm eq}\,,
\end{equation}
and 
\begin{equation}
\langle S \rangle = q(T,t)S_{\rm eq}\,,
\end{equation}
after we have substituted $\langle c \rangle$ and $\langle S \rangle$ in equation 
(16) and regrouped terms we have
\begin{equation}
\left(\frac{\partial S_{\rm eq} }{\partial T}-\frac{N}{T}c_{\rm eq}\right)q
=-S_{\rm eq}\frac{\partial q}{\partial T}\,.
\end{equation}
The left term of this equation is zero while the right term is different of zero
when the system is not at the equilibrium. So that we can write a new relation for the
entropy and the specific heat. It follows as
\begin{equation}
\frac{\partial \langle S \rangle}{\partial T}-S_{\rm eq}\frac{\partial q}{\partial T}
=\frac{N}{T}\langle c \rangle \,.
\end{equation}
At the equilibrium $\partial q/\partial T$ vanishes and we
recover equation (16). In order to check this expression
in Figure 4 we compare the specific heat calculated from this equation, and from
the values of the entropy, with the specific heat calculated from equation (9). 
The derivatives are approximated with finite differences.  
We can see that both agree very well in the long time regime.

\section{Conclusion}

We have studied the relaxation process of the specific heat and the entropy for 
the three-dimensional Coulomb glass at very low temperatures and in thermal equilibrium.  
The long time relaxation regime follows a stretched exponential function, with the 
exponent $\beta<1$ increasing with the temperature, which is an indication of a broad 
distribution of relaxation times in the system. 
From these results, we could expect a dynamical transition at some temperature 
$T^*$ (above of the temperatures studied here) where $\beta=1$. The exponential relaxation 
process would be recovered for $T>T^*$. In reference \cite{WS97} it was found $T^*>0.03$ 
for the two-dimensional Coulomb glass although the effects of distance-dependent 
transition probabilities were not taken into account. Moreover, the relaxation 
functions found here in the thermal equilibrium are very different from the power 
law relaxations found when the system is out of equilibrium \cite{PO98}. Nevertheless,
in out of equilibrium experiments stretched exponential relaxation functions are 
found \cite{VO00}. The relaxation time diverges as an Arrhenius law when $T\rightarrow 0$.  
Finally, we have found a relation between the specific heat and the entropy in the long
time regime. 

\begin{acknowledgement}
This work was supported by the European TMR Network-Fractals (Contract No. FMRXCT980183).
A. D\'\i az-S\'anchez acknowledges support from a Postdoctoral Grant from the European
TMR Network-Fractals. We are indebted to M.~Ortu\~no, A.~M\"obius, M.~Pollak, and 
A. Coniglio for stimulating discussions. 
\end{acknowledgement}

\begin{figure}
\epsfxsize=\hsize
\begin{center}
\leavevmode
\epsfbox{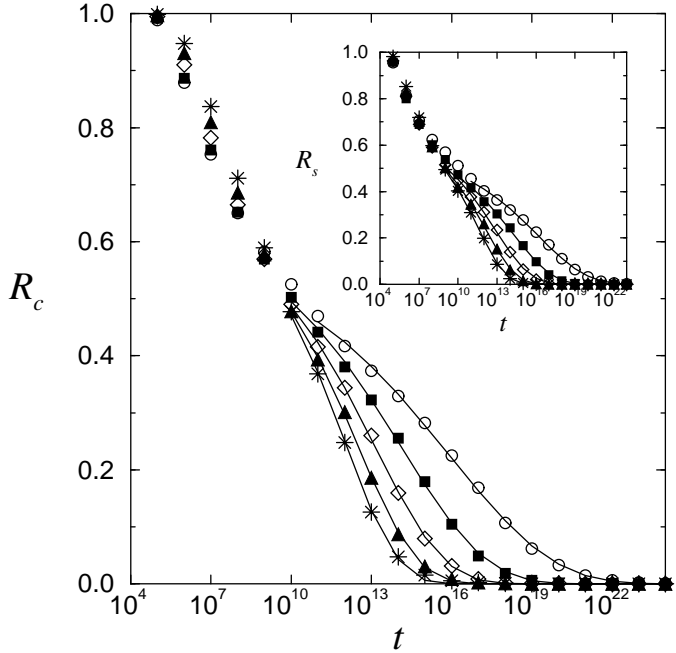}
\end{center}
\caption{Relaxation functions of the specific heat as a function of $t$, for the
temperatures $T= 0.006$ ($\circ$), 0.008 ($\blacksquare$), 0.01 ($\diamond$), 
0.012($\blacktriangle$) and 0.014 ($*$). Inset the relaxation functions of the entropy
for the same temperatures. The solid lines correspond to fits with the function
$f(t)=f_0\exp\left[-(t/\tau)^\beta\right]$.}
\end{figure}

\begin{figure}
\epsfxsize=\hsize
\begin{center}
\leavevmode
\epsfbox{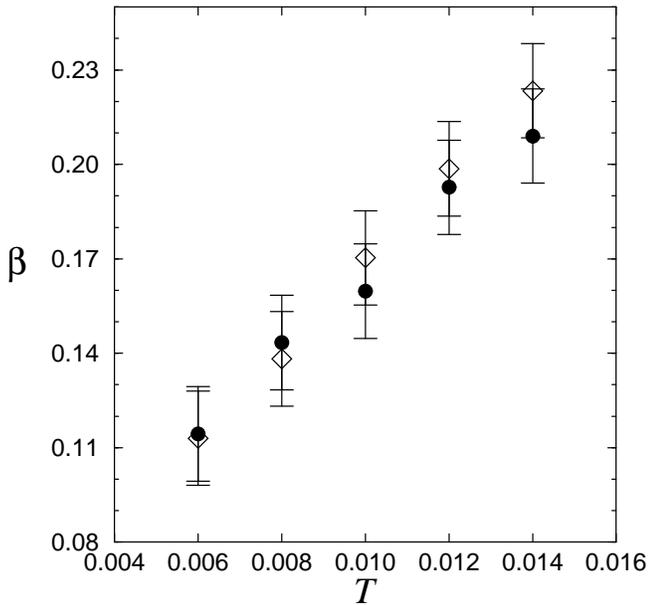}
\end{center}
\caption{Parameter $\beta$ as a function of the temperature, obtained by fitting 
$R_c$ ($\bullet$) and $R_s$ ($\diamond$) with the function 
$f(t)=f_0\exp\left[-(t/\tau)^\beta\right]$.}
\end{figure}

\begin{figure}
\epsfxsize=\hsize
\begin{center}
\leavevmode
\epsfbox{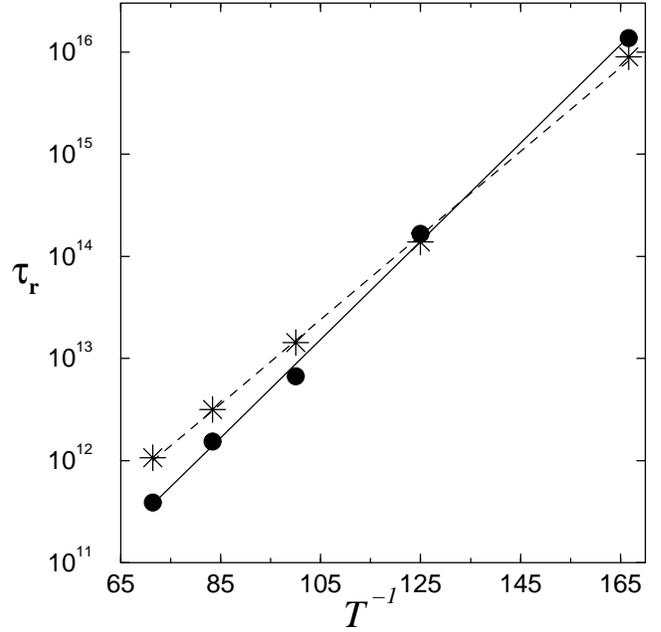}
\end{center}
\caption{Relaxation time $\tau_r$ as a function of the inverse of the temperature,
obtained by fitting $R_c$ ($\bullet$) and $R_s$ ($*$) with the function
$f(t)=f_0\exp\left[-(t/\tau_r)^\beta\right]$. The lines are fits with the function
$\tau_r(T)=a\exp[T_0/T]$. The values of $\tau_r$ correspond to the
values obtained from the fit in Figure 1.} 
\end{figure}

\begin{figure}
\epsfxsize=\hsize
\begin{center}
\leavevmode
\epsfbox{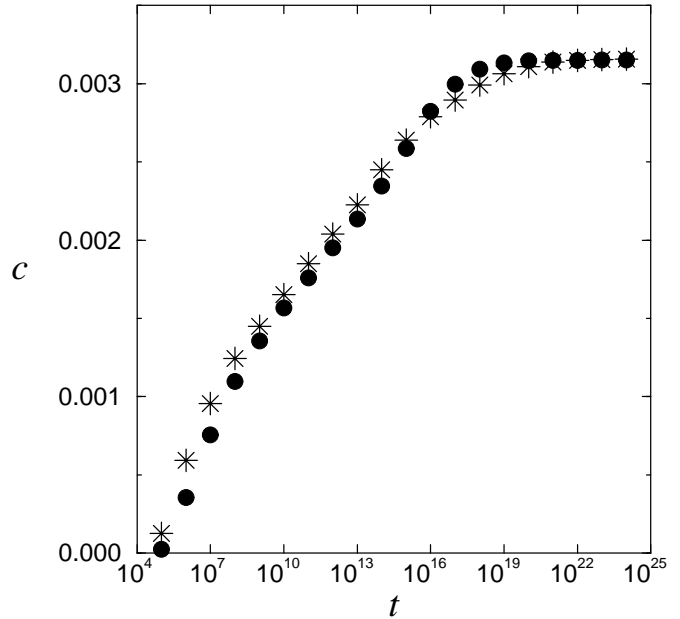}
\end{center}
\caption{Specific heat as a function of time obtained from equation (20) and the
values of the entropy ($*$) and from equation (9) ($\bullet$), for $T=0.008$.}
\end{figure}

\end{document}